\documentclass[conference]{IEEEtran}
\IEEEoverridecommandlockouts
\usepackage{cite}
\usepackage{amsmath,amssymb,amsfonts}
\usepackage{algorithmic}
\usepackage{graphicx}
\usepackage{textcomp}
\usepackage{enumerate}
\usepackage{makecell}
\usepackage{xcolor}
\usepackage[normalem]{ulem}

\makeatletter
\def\ps@headings{%
\def\@oddhead{\mbox{}\scriptsize\rightmark \hfil \thepage}%
\def\@evenhead{\scriptsize\thepage \hfil \leftmark\mbox{}}%
\def\@oddfoot{}%
\def\@evenfoot{}}
\makeatother
\def\BibTeX{{\rm B\kern-.05em{\sc i\kern-.025em b}\kern-.08em
    T\kern-.1667em\lower.7ex\hbox{E}\kern-.125emX}}

\usepackage{url}

\def\BibTeX{{\rm B\kern-.05em{\sc i\kern-.025em b}\kern-.08em
    T\kern-.1667em\lower.7ex\hbox{E}\kern-.125emX}}

\newcommand{\delete}[1]{}

\begin{document}

\title{BOSS: A Blockchain Off-State Sharing System\\
}

\author{
\IEEEauthorblockN{Shan Wang\IEEEauthorrefmark{1}\IEEEauthorrefmark{3},
Ming Yang\IEEEauthorrefmark{1},
Tingjian Ge\IEEEauthorrefmark{3},
Yan Luo\IEEEauthorrefmark{3},
Xinwen Fu\IEEEauthorrefmark{3}
}
\IEEEauthorblockA{\IEEEauthorrefmark{1}
Southeast University. Email: shanwangsec@gmail.com, yangming2002@seu.edu.cn}
\IEEEauthorblockA{\IEEEauthorrefmark{3}University of Massachusetts Lowell. Email:\{Yan\_Luo, Xinwen\_Fu\}@uml.edu, ge@cs.uml.edu}
}

\maketitle

\begin{abstract}
Blockchain has been applied to data sharing to ensure the integrity of data and chain of custody.
Sharing big data such as large biomedical data files is a challenge to blockchain systems since the ledger is not designed to maintain big files, access control is an issue, and users may be dishonest.
We call big data such as big files stored outside of a ledger that includes the blockchain and world state at a blockchain node as ``off-state" and propose an off-state sharing protocol for a blockchain system to share big data between pairs of nodes. In our protocol, only encrypted files are transferred. The cryptographic key is stored in the world state in a secure way and can be accessed only by authorized parties.
A receiver has to request the corresponding cryptographic key from the sender to decrypt such encrypted files. All requests are run through transactions to establish reliable chain of custody. 
We design and implement a prototypical blockchain off-state sharing system, BOSS, with Hyperledger Fabric. Extensive experiments were performed to validate the feasibility and performance of BOSS.
\end{abstract}

\begin{IEEEkeywords}
Blockchain, Big File Sharing, Off-state, Hyperledger Fabric
\end{IEEEkeywords}

\section{Introduction}

A blockchain system can build trust in the data that it maintains without a centralized authority. Data in conventional blockchain systems is often stored in a ledger, which includes a world state and a blockchain. The world state stores the current system state such as the user cryptocurrency balance in Bitcoin \cite{antonopoulos2014mastering} and the blockchain saves all transaction history, which contains operations on the world state and/or the data used to update the world state. The whole transaction history produces the current state values and can work as auditing evidence. The ledger is often synchronized across all nodes.

Blockchain has been applied to a variety of data sharing applications such as healthcare.
A blockchain data sharing system can prevent information tampering for the purpose of auditing and establish the chain of custody, which refers to the documentation that records the sequence of custody, transfer, and other operations on the data.
In these systems, shared data is stored in the world state database. Users set or get shared data through proposing transactions to the blockchain system.

The existing blockchain frameworks cannot be directly applied to big file sharing.
Big files can be scientific and biomedical data being collected and processed. Sharing such data allows independent verification of published scientific results and enhances opportunities for new discoveries. With concerns of intellectual property (IP) theft and industrial espionage, we desire secure and trustworthy big file sharing systems that record the chain of custody of the big files. However, given the storage design of the ledger, it is impractical to store big files in the world state and to use transactions to carry big files. A blockchain system often maintains the same ledger across all nodes. However, because of privacy, IP and storage concerns, owners may not want to share those big files across all nodes. 

A token based data sharing blockchain system has been briefly discussed in related work \cite{zhang2018fhirchain}. Shared data is stored outside the blockchain system at a data center.
This system has the following issues: (i) It requires a centralized and trusted data center. (ii) No actual protocols using tokens are documented. (iii) The data center based approach is not flexible. In scenarios such as file sharing in biomedical fields, researchers may want to share data between each other's computers directly. There is no detail of the token based sharing protocol in \cite{zhang2018fhirchain}.

In this paper, we propose a novel big file sharing model. Our major contributions can be summarized as follows.

\begin{enumerate}
\item We introduce a new concept ``off-state" to a blockchain system. A blockchain node may maintain a separate storage from the ledger to store off-states such as big files, which are outside of the ledger, particularly the world state. Off-states do not need to be synchronized across all nodes. Smart contracts manage sharing of off-states such as big files.

\item To ensure the integrity and chain of custody of off-states, we propose a novel off-state sharing protocol. Off-sates are encrypted and signed before sharing. The cryptographic key is stored in the world state database and users have to request the key through transactions in order to decrypt the transferred files. Mechanisms such as private data collection (PDC) in Hyperledger Fabric \cite{androulaki2018hyperledger} can be used to ensure that the cryptographic key is kept private and not publicly disclosed. These mechanisms also defeat dishonest users.

\item We implement a prototypical blockchain off-state sharing system (BOSS) with Hyperleger Fabric running our off-state data sharing protocol. Extensive experiments are performed to evaluate BOSS' feasibility and performance such as latency of the big file transfer between pairs of blockchain nodes and parallel file transfer.

\end{enumerate}

The rest of this paper is organized as follows. Section \ref{sec::background} introduces the background knowledge. Section \ref{sec::system Model} introduces the off-state sharing system model. The off-state sharing protocol is presented in Section \ref{sec::Integration_Scheme}. We evaluate the prototypical off-state sharing system, BOSS, in Section \ref{sec::Evaluations}. We discuss a few related topics such as the execution timeout of chaincode (i.e. smart contract in Hyperledger Fabric) in Section \ref{sec::discussion}. Section \ref{sec::related_work} presents the related work. We conclude this paper in Section \ref{sec::Conclusion}.

\section{Background}
\label{sec::background}

In this section, we introduce the data in blockchain systems, storage limitations of a ledger, and Hyperledger Fabric, which is a permissioned blockchain framework. For brevity, we will call Hyperledger Fabric as Fabric in the rest of the paper.

\subsection{Data in Conventional Blockchain Systems}
\label{sec::world_state}

A blockchain system is essentially a state machine replicated across different nodes \cite{buterin2014next}. The state machine can be formally defined in Formula (\ref{eqn::enctypt_aes}).
\begin{eqnarray}
APPLY(S, Tx) \rightarrow S' or ERROR,
\label{eqn::enctypt_aes}
\end{eqnarray}
where the state transition function $APPLY(.)$ takes a state $S$ and a transaction $Tx$ and outputs a new state $S'$ or an error. 
A ledger includes a world state and a blockchain. 
The world state stores the current state. The blockchain stores all historical transactions.
In Bitcoin \cite{antonopoulos2014mastering}, the world state is the user cryptocurrency balance recorded in the UTXO (unspent transaction output) field of transactions.
State transitions are cryptocurrency transfers. 
In Ethereum \cite{buterin2014next}, the world state is made up of ``accounts". An account contains the cryptocurrency balance, code and data in smart contracts and others. State transitions are direct transfers of values and information between accounts controlled by smart contracts. In Fabric \cite{androulaki2018hyperledger}, each peer node maintains a world state database. A transaction records the operation of invoking a function in smart contract, called chaincode in Fabric, which may work on the world state. \looseness=-1

\subsection{Storage Limitations of Conventional Blockchain Systems}
\label{sec::Storage_Limitations}

In current blockchain frameworks, the data size and type in ledgers are limited. 
We now show the limitations in three popular blockchain frameworks, Bitcoin, Ethereum and Fabric. 

\subsubsection{\textbf{World State Storage Limitation}}

In Bitcoin, the world state UTXOs are just values in transactions indicating the Bitcoin balance. It's impossible to store large files in UTXOs.

In Ethereum, the world state consists of user accounts and contract accounts. A user account only records the Ether balance of the user. The contract account can store pairs of keys and values of 32 bytes used by the smart contract \cite{EthereumBuildersGuideline}.
Ethereum is not designed to store big files.

In Fabric, the world state is stored in a database, i.e. LevelDB or CouchDB. LevelDB can only store simple key-value pairs. CouchDB can store data in the JSON format, but limits the maximum document body size to 8M bytes \cite{couchDB_Doc}. Fabric cannot store big files either.

\subsubsection{\textbf{Transaction and Block Size Limitation}}

In Bitcoin, to prevent attacks like DoS attacks and memory exhaustion attacks, it restricts the block size to 1MB and transactions greater than 100 KB are considered non-standard \cite{BitcoinWeakness,BitcoinForum}.

In Ethereum, the {\em gas} consumed by a transaction is related to the size of the data stored in the world state. It is quite expensive to store big data. According to the yellow paper \cite{EthYellowPaper}, it requires 16 gas units for storing every non-zero byte of data for a transaction. Storing 1MB ($1024 \times 1024 = 1048576 \ bytes$) data requires 16,777,216 gas units in addition to other fees. Each block has a block gas limit, which limits the block size too.
The current block gas limit is 12,491,593 gas units/block \cite{GasLimit}.
It means a block can maximally store about 0.74MB data ($12491593 \div 16777216 \approx 0.74$).\looseness=-1

In Fabric, the preferred maximum size of each transaction is 512KB while the absolute maximum transaction size is 99MB. Nodes in Fabric adopt the {\em grpc} communication protocol. The maximum inbound grpc message size is 100MB \cite{HyperledgerJavazAPI}. Large transactions will decrease the system performance in terms of transaction throughput \cite{gorenflo2020fastfabric}. 
\looseness=-1

\subsection{Hyperledger Fabric}
\label{sec::hyperledger_fabric}

Fabric \cite{androulaki2018hyperledger} is a popular permissioned blockchain framework.
It decouples the tasks of a node in blockchain systems such as Ethereum \cite{buterin2014next} into three types of nodes, i.e. peers, orderers and clients in charge of different tasks. Peers maintain ledgers and smart contracts.

\begin{figure}
\centering
\includegraphics[width=1\columnwidth]{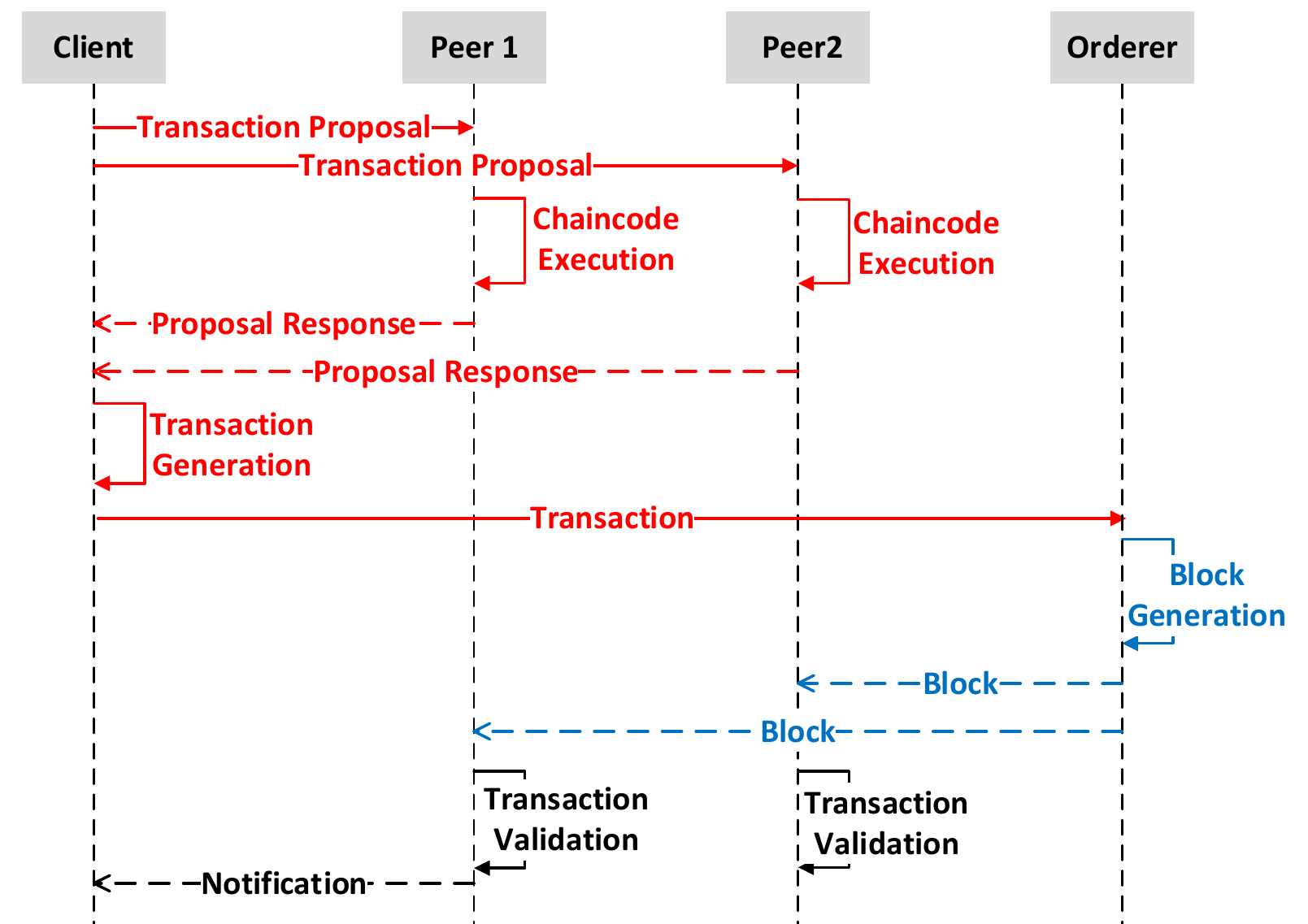}
\caption{Three-phase transaction workflow in Fabric
}
\label{fig:txworkflow} 
\end{figure}

\textit{Transaction Workflow:}
\label{subsubsec::FabricTransactionWorkflow}
Fabric adopts a three-phase ``execute-order-validate'' transaction workflow as shown in Fig. \ref{fig:txworkflow}: (1) A client/user proposes a transaction {\em proposal} to the endorsers, which are specific peers and specified by the endorsement policy; (2) The endorsers execute the chaincode, sign the execution results, and return the results with corresponding signatures to the client as proposal responses; (3) If the returned execution results from different endorsers are the same, the client constructs a transaction, which contains both the transaction proposal, proposal response and a list of signatures (endorsements). The client
sends the transaction to the orderer nodes. (4) The orderer nodes collect multiple transactions within a time period, bundle them into a new block, and distribute the new block to all peers including both endorsers and non-endorsers. (5) All peers validate the transactions in the received new block. If all transactions are valid, each peer will update the world state according to execution results. After validating all transactions, each peer appends the new block to its local blockchain.

Three features of Hyperledger Fabric enable it suitable to implement our blockchain big file sharing system.

\subsubsection{Fine-grained data isolation}
The Fabric introduces a fine-grained data isolation mechnism, i.e., private data collection (PDC). PDC data is sensitive and shared by only a subset of peer nodes. A PDC is stored in the world state. Only PDC member peers can store the original data, while PDC non-member peers can only store the data hashes.

\subsubsection{Customizable smart contracts}
Smart contracts, also called chaincode in Fabric, do not need to be identical at all peers as long as the execution results across different endorser peers are the same. 
Therefore, different peers can perform different roles and tasks.

\subsubsection{Multi-level endorsement policy}
There are multiple levels of endorsement policies in Fabric. An endorsement policy stipulates which peers need perform as endorsers to endorse a transaction. Endorsing involves executing smart contracts and signing the execution results.
Each smart contract has a default chaincode-level endorsement policy that manages all public data and PDC data in the world state. A collection-level endorsement policy can be customized to specifically manage the PDC data. 

\section{Off-State Data Sharing System Model}
\label{sec::system Model}

In this section, we first investigate the existing blockchain-based big file sharing systems and their drawbacks. We then present the challenges in big file sharing through blockchain systems, and finally propose a new big file sharing model through blockchain.

\subsection{State-of-the-art Blockchain Based Big File Sharing Systems}
\label{sec::Existing_System_Models_Analysis}

\begin{figure}
\centering
\includegraphics[width=0.85\columnwidth]{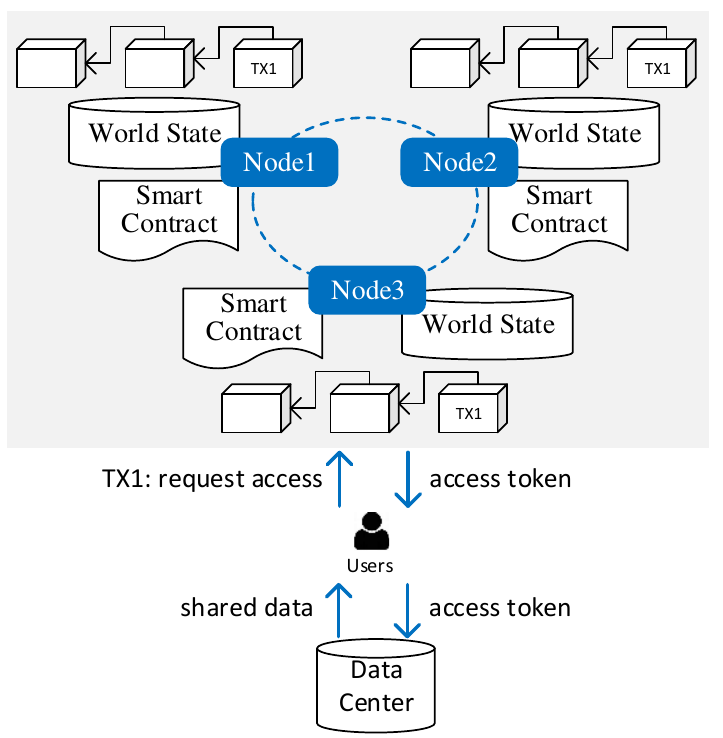}
\caption{Token based blockchain data sharing system
}
\label{fig:exist_system_model2} 
\end{figure}

Fig. \ref{fig:exist_system_model2} shows a blockchain-based data sharing model utilizing access tokens \cite{zhang2018fhirchain}. 
A user proposes a transaction to request a token for data access. The transaction proposal triggers particular functions in the smart contract, which returns an access token to the user. 
The user then uses the access token to request the shared data from a data center, which is outside of the blockchain system. This data sharing model may be used for big file sharing.

The token based file sharing blockchain model has the following issues: (i) It requires a centralized and trusted data provider. 
The data center is controlled by the data provider.
The Blockchain system has no control over the actual data access and cannot know the status of data sharing, e.g. whether the data is actually shared. 
The integrity of chain of custody cannot be ensured and audited,
(ii) There is no detailed presentation of the actual protocol in \cite{zhang2018fhirchain}.
(iii) The data center based approach is not flexible. In scenarios such as file sharing in biomedical fields, data owners may want to share data between their workstations. 

\subsection{Design Challenges}

We identify three challenges for a blockchain-based big file sharing system.

\subsubsection{Challenge 1---Storage Space Limitation}

Big files refer to files of large size, such as videos and biomedical data files.
Due to the storage limitations of ledgers as discussed in Section \ref{sec::Storage_Limitations}, it is impractical to store big files in ledgers or pack big files in transactions. 

\subsubsection{Challenge 2---Privacy Requirement}
\label{sec::challenge2}

Big files may be sensitive. The owner may not want to share with everyone and access control will be preferred.
However, in a conventional blockchain system, all nodes maintain the same data. Therefore, the blockchain system cannot be directly used for big file sharing.

\subsubsection{Challenge 3---Security Requirement}
\label{sec::challenge3}
In big file sharing, the receiver of the big files may be dishonest and may dishonestly deny she/he received the files from the owner.

\subsection{Off-State Sharing System Model}
\label{sec::off-stateConcept}

We introduce the concept of ``off-state", which is stored at blockchain nodes, but outside of ledgers, particularly the world state. For example, the off-state can be big files.
Therefore, we generalize big file sharing as off-state sharing. 
Off-states can be inconsistent across blockchain nodes and shared between pairs of nodes. Smart contracts at blockchain nodes can interact with off-states and perform the sharing operation. 

\begin{figure}
\centering
\includegraphics[width=1\columnwidth]{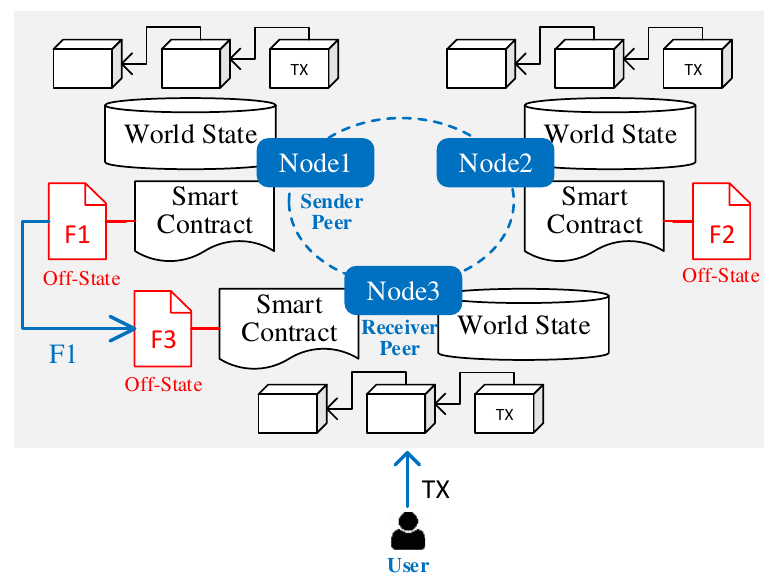}
\caption{Blockchain-based off-state sharing model
}
\label{fig:system_model} 
\end{figure}

Fig. \ref{fig:system_model} shows our novel blockchain system model that can address the three challenges above for off-state data sharing. Each node maintains a world state database and an off-state storage space when needed. Nodes maintain their off-state data individually. Users propose transactions to trigger the smart contracts to transfer the files. The smart contracts at the sender node transfers off-state data to the receiver node through a file transfer protocol, not through transactions. 

To summarize, we address the three challenges as follows.
\begin{itemize}
\item 
To address Challenge 1---storage space limitation, we introduce the concept of `off-state" and save big files off the ledger. 

\item To address Challenge 2---privacy requirement, our protocol allows off-states to be shared between pairs of nodes, not across all blockchain nodes. This feature saves storage space at nodes that do not need off-state data.

\item To address Challenge 3---security requirement, our protocol establishes the chain of custody of the off-state data and records auditing evidence in the blockchain.
\end{itemize}
We will present the detailed off-state data sharing protocol in Section \ref{sec::Integration_Scheme}.


\section{Off-State Sharing Protocol}
\label{sec::Integration_Scheme}

Our protocol has two stages. In Stage 1, an owner prepares the files to be shared with users and uploads the files to sender peer nodes. 
In Stage 2, to establish the chain of custody, a sender peer transfers an encrypted file to a receiver peer and stores the symmetric encryption/decryption key (called encryption key for brevity in the rest of the paper) in the world state. A user has to request the encryption key through a transaction to get the key, decrypt the encrypted file and get the original file. These transactions establish the chain of custody.
In this section, we first present the two stages and then perform the security analysis. We use Fabric to present particular protocol details and discuss the use of other blockchain frameworks for the protocol in Section \ref{sec::discussion}.

\begin{figure}
\centering
\includegraphics[width=1\columnwidth]{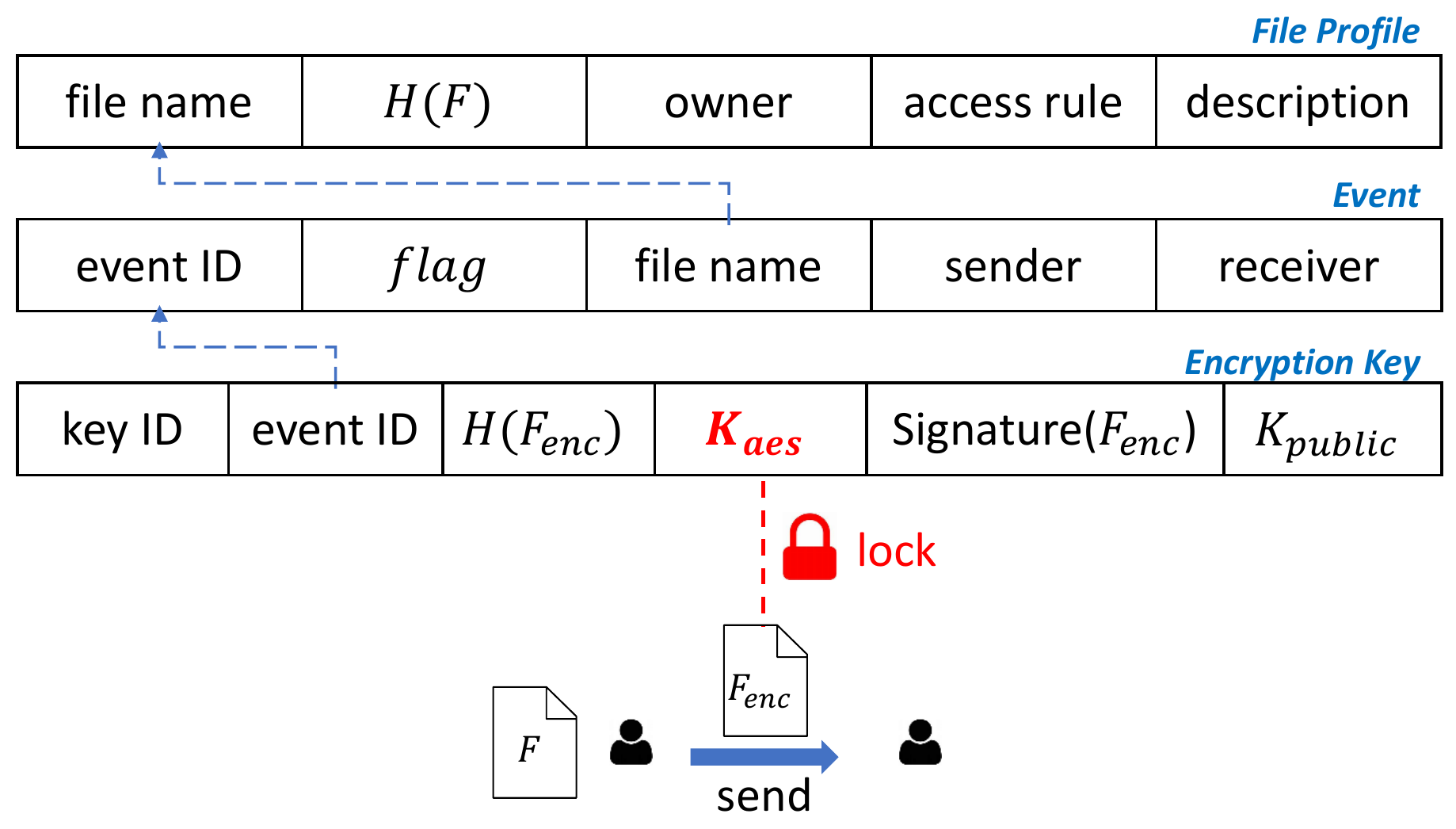}
\caption{Cryptographic lock mechanism for off-state
}
\label{fig::Module2DataPlane} 
\end{figure}

\begin{table}
\centering
\caption{Summary of Main symbols.}
\begin{tabular}{l|l}
\hline
Symbol               & Description                                             \\ \hline
$F$                    & File to be transferred                             \\
$H(F)$                 & Hash of F                                           \\
$K_{aes}$            & Symmetric key        \\
$F_{enc}$            & File encrypted using $K_{aes}$                    \\
$H(F_{enc})$         & Hash of $F_{enc}$                                  \\
$K_{private}$        & Private key to sign $F_{enc}$ \\
$K_{public}$         & Public key to verify digital signature\\
$Signature(F_{enc})$ & Digital signature of $F_{enc}$                     \\ \hline
\end{tabular}
\label{table::symbols}
\end{table}

\subsection{Stage 1: Preparing Off-State Data}

A data owner uploads big files to the off-state storage of her/his own blockchain node. The user also proposes a transaction---``Tx: file upload"---to record the profile of an uploaded file into the world state.
The file profile in the world state is public for searching.
Fig. \ref{fig::Module2DataPlane} illustrates the \texttt{File Profile} entry in the world state, \texttt{ $\langle$file\_name, $H(F)$, owner, access\_rule, description$\rangle$}, where $H(F)$ is the hash of the file. $H(F)$ can be used to verify the integrity of a received file. 

\subsection{Stage 2: Sharing Off-State Data}
\label{Sharing_Off_State_Data}

\begin{figure}
\centering
\includegraphics[width=1\columnwidth]{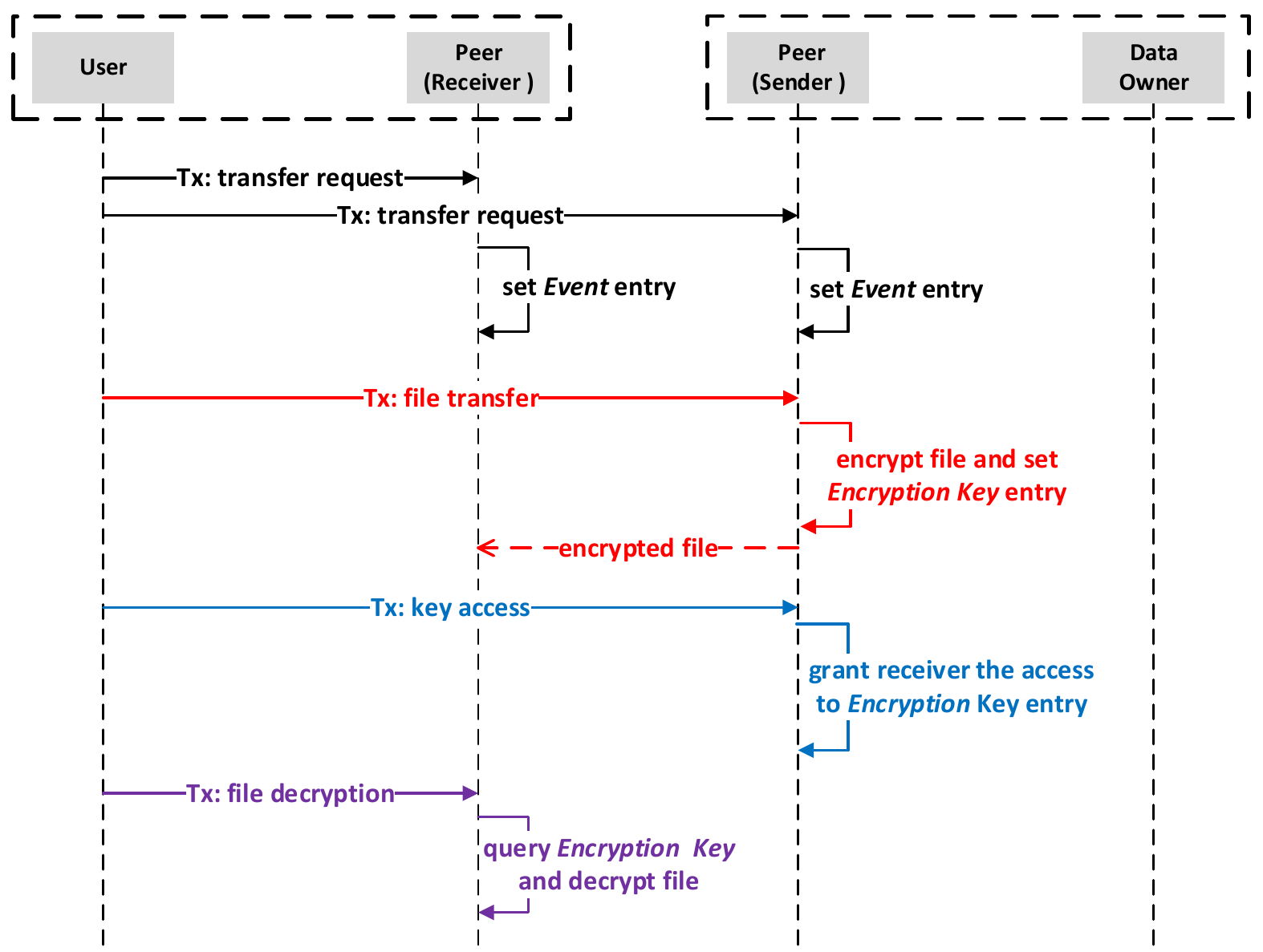}
\caption{Four phases in the off-state sharing protocol
}
\label{fig::Module2} 
\end{figure}

Fig. \ref{fig::Module2} shows the workflow of sharing off-state data, e.g. a big file, in Fabric. It involves four transactions. These four transactions follow the same transaction lifecycle as introduced in Section \ref{subsubsec::FabricTransactionWorkflow}. However endorsers in each transaction can be different according to the specific endorsement policy setting. Fig. \ref{fig::Module2} shows which peers perform as endorsers to execute the chaincode in each transaction lifecycle.

\subsubsection{Phase 1: File Transfer Request}
A user shall request the transfer of a particular file stored at the sender peer and the request shall pass the access control.
In Fabric, a user can propose a transaction to request a particular file, and the access control is governed by a specific endorsement policy. For example, the endorsement policy can be set as AND(Peer.Sender, Peer.Receiver), which stipulates both sender peer and receiver peer should agree on the file transfer.
An endorser checks if access rules in the corresponding \texttt{File Profile} are satisfied through smart contracts, and records the results in an \texttt{Event} entry of the world state. 
An \texttt{Event} entry is in the form of \texttt{ $\langle$event\_ID, flag, file\_name, sender, receiver$\rangle$} as shown in Fig. \ref{fig::Module2DataPlane}, where the $flag$ is a Boolean value indicating if the endorser permits the file transfer. 
According to the transaction workflow in Fabric introduced in Section \ref{subsubsec::FabricTransactionWorkflow}, when results (i.e. \texttt{Event} entry) returned from different endorsers are the same, the transaction can be generated, processed and recorded in a blockchain. The same \texttt{Event} entries mean the sender and receiver reach a consensus, either denial of or agreement on the file transfer. We call the corresponding transaction ``Tx: transfer request".

\subsubsection{Phase 2: Encrypted File Transfer}
After the file transfer request is approved, the user proposes another transaction---``Tx: file transfer"---to initiate the file transfer. In Fabric, the endorsement policy for ``Tx: file transfer" can be set as AND(Peer.Sender), which means only the sender peer works as an endorser.
The sender peer/endorser first checks $flag$ in the corresponding \texttt{Event} entry. The file transfer continues only if $flag$ is $True$.

The sender peer can then encrypt and sign the file, and send the encrypted file to the receiver peer via a file transfer protocol such as \textit{SFTP}. 
The receiver peer stores the encrypted file in its off-state storage.
In Fabric, the encryption key and related information of the encrypted file are put in the world state as an \texttt{Encryption Key} entry which is a PDC. PDC can guarantee that only PDC members can see the actual encryption key and others only know its hash. In this phase, only the sender peer is the PDC member. The \texttt{Encryption Key} entry is in the form of \texttt{$\langle$key\_ID, event\_ID $H(F_{enc})$, $K_{aes}$, $Signature(F_{enc})$, $K_{public}$ $\rangle$}, where $K_{aes}$ is the key to decrypt the encrypted file.


\subsubsection{Phase 3: Key Retrieval}
A user first checks if the receiver peer has received the transferred file. Only after the receiver peer has received the file does the user propose a transaction---``Tx: key access"---to request $K_{aes}$ from the sender peer to decrypt the encrypted file $F_{enc}$. Otherwise, the user does nothing. In Fabric, a user can send a query request to the receiver peer to check if $F_{enc}$ exists. The query request will not generate a transaction. Recall only the sender peer stores the original $K_{aes}$ and other nodes store the hash. The endorsement policy for ``Tx: key access" is set as AND(Peer.Sender) so that the sender peer performs as an endorser and adds the receiver peer to the PDC member list. Therefore, the receiver peer will be able to retrieve $K_{aes}$.

\subsubsection{Phase 4: File Decryption}
Finally, the user proposes the last transaction ---``Tx: file decryption"---to query $K_{aes}$ to decrypt $F_{enc}$. In Fabric, the endorsement policy can be set as AND(Peer.Receiver) so that the receiver peer performs as an endorser, which can access $F_{enc}$ in its off-state storage using smart contract. The receiver peer first queries the \texttt{Encryption Key} entry in the world state for $K_{aes}$ and then decrypt $F_{enc}$.

\subsection{Security Analysis}
\label{sec::stage3}


We now show that the mechanisms in our off-state sharing protocol can ensure the integrity of chain of custody of the off-state data. That is, the user cannot deny that she/he has received the requested off-state data.
We assume the underlying blockchain infrastructure, i.e. all peers, is secure, and smart contracts execute as designed.
The assumption is reasonable since the peer administrators can adjust the user privilege to meet the need. \looseness=-1

An auditor can find out if a user has received a transferred file by searching for two transactions ``Tx: key access" and ``Tx: file decryption" related to the particular file from the blockchain. In our protocol, the receiver peer has to obtain $K_{aes}$ in order to decrypt the encrypted file $F_{enc}$, which the sender sends.
$K_{aes}$ is stored in the world state of the ledger, which guarantees the integrity of the data because of the underlying blockchain framework. When a user proposes the transaction ``Tx: key access", the receiver peer will receive $K_{aes}$. 
Once the receiver peer obtains $K_{aes}$, the user has to propose the transaction ``Tx: file decryption" to get decrypted file $F$. It can be observed that if the receiver peer does not receive $F_{enc}$, the user will not propose ``Tx: key access". If the receiver peer receives $F_{enc}$ but the user does not propose ``Tx: key access" to request $K_{aes}$, the receiver peer can not get the original file $F$.

An auditor can also find out if a sender peer has transferred a file by searching for ``Tx: file transfer" related to the particular file from the blockchain. In our protocol, ``Tx: file transfer" will trigger smart contracts to transfer a file in the off-state storage and stores \texttt{Encryption Key} in the world state. The integrity of ``Tx: file transfer" can be guaranteed by the underlying blockchain framework.

A user may wrongfully claim she/he receives a wrong file $f$. The transaction history maintained by the blockchain system actually provides proofs to dispute the claim. 
``Tx: file upload" and ``Tx: transfer request" can provide proofs to verify the integrity of the transferred file. In our protocol, a file owner records \texttt{File Profile} in the world state by proposing ``Tx: file upload". The sender peer and receiver peer should agree on a file sharing event which is recorded in the \texttt{Event} entry by ``Tx: transfer request". The file name in \texttt{Event} is related to a particular \texttt{File Profile}. ``Tx: transfer request" in blockchain can prove the sender and receiver peers have agreed on transferring a specific file and $H(F)$ in \texttt{File Profile} can be used to verify if $f$ matches with $H(F)$.


\section{Evaluation}
\label{sec::Evaluations}

We have implemented a prototypical off-state sharing system, {\em BOSS}, based on Hyperledger Fabric, and evaluate its feasibility and performance in this section.

\subsection{Experiment Setup}
\label{sec::systemFramework}

\begin{figure}
\centering
\includegraphics[width=1\columnwidth]{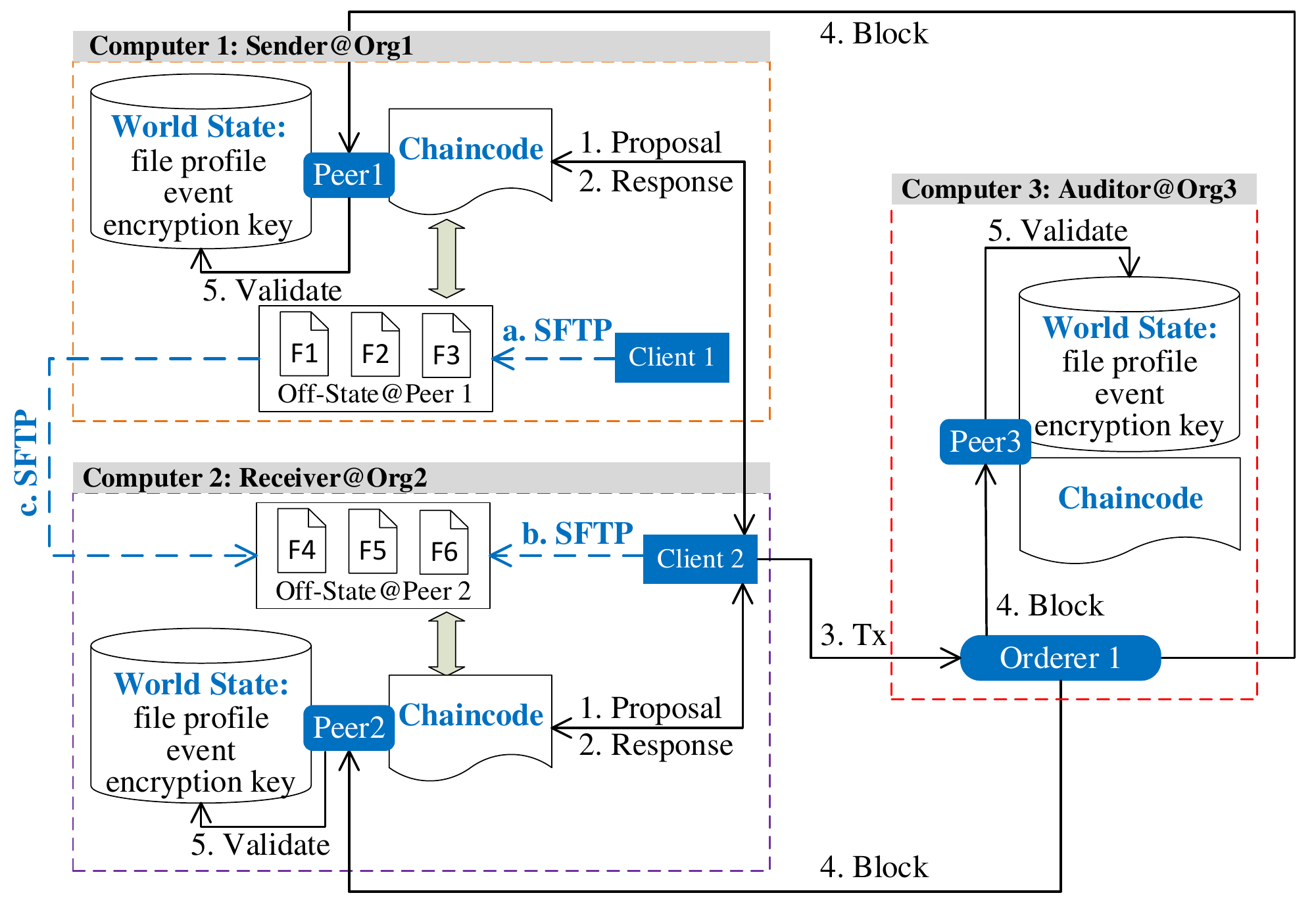}
\caption{Prototypical off-state sharing system (BOSS) based on Hyperledger Fabric
}
\label{fig::framework} 
\end{figure}

Fig. \ref{fig::framework} shows the prototypical BOSS, which has three organizations, i.e. org1, org2 and org3. Org1 and org2 are the sender and the receiver, and each party contributes one peer node and one client (user/owner) node. Org3 is an auditing organization, and contributes one peer node and one orderer node. We implement our off-state sharing protocol using chaincode of Fabric and adopt \textit{SFTP} \cite{sftp_Wikipedia} to perform file transfer.

It can be observed that in our system, peer nodes are different from the conventional Fabric peer nodes. We integrate an off-state storage and the \textit{SFTP} service into peer nodes of interest. The chaincode can transfer and retrieve files from the local off-state storage. 

We implement the prototypical system and deploy it across three physical computers. Table \ref{table::setup} lists the configuration of the computers for sender, receiver and auditor nodes.
The sender computer and the receiver computer are located in different buildings on a university campus. All Fabric nodes run in Docker containers, which form the Fabric network. We use \textit{Golang} to develop the chaincode and use \textit{Node.js} to develop the client application that a user uses to propose transactions to the peer nodes.

\begin{table}[ht]
\centering
\caption{Blockchain Node Information in Prototypical BOSS}
\setlength{\tabcolsep}{0.3mm}{
\begin{tabular}{c|c|c|c|c|c}
\hline
Computer & Role                  & OS                                                 & Memory & Disk & CPU                                                                                            \\ \hline
1      & Sender   & \begin{tabular}[c]{@{}c@{}}Ubuntu\\ 18.04\end{tabular} & 64GB   & 1TB  & \begin{tabular}[c]{@{}c@{}}Intel Xeon(R) Gold\\ 6128 CPU@3.40GHz\\ ×24 Processor\end{tabular}  \\ \hline
2      & Receiver     & \begin{tabular}[c]{@{}c@{}}Ubuntu\\ 18.04\end{tabular} & 32GB   & 1TB  & \begin{tabular}[c]{@{}c@{}}Intel Core i7-6700\\ CPU@3.40GHz\\ × 8 Processor\end{tabular}       \\ \hline
3      & Auditor  & \begin{tabular}[c]{@{}c@{}}Ubuntu\\ 16.04\end{tabular} & 64GB   & 1TB  & \begin{tabular}[c]{@{}c@{}}Intel Xeon(R) Gold\\ 6128 CPU@3.40GHz\\ × 49 Processor\end{tabular} \\ \hline
\end{tabular}}
\label{table::setup}
\end{table}

\subsection{Performance and Feasibility}

\begin{figure}[ht]
\centering
\includegraphics[width=1\columnwidth]{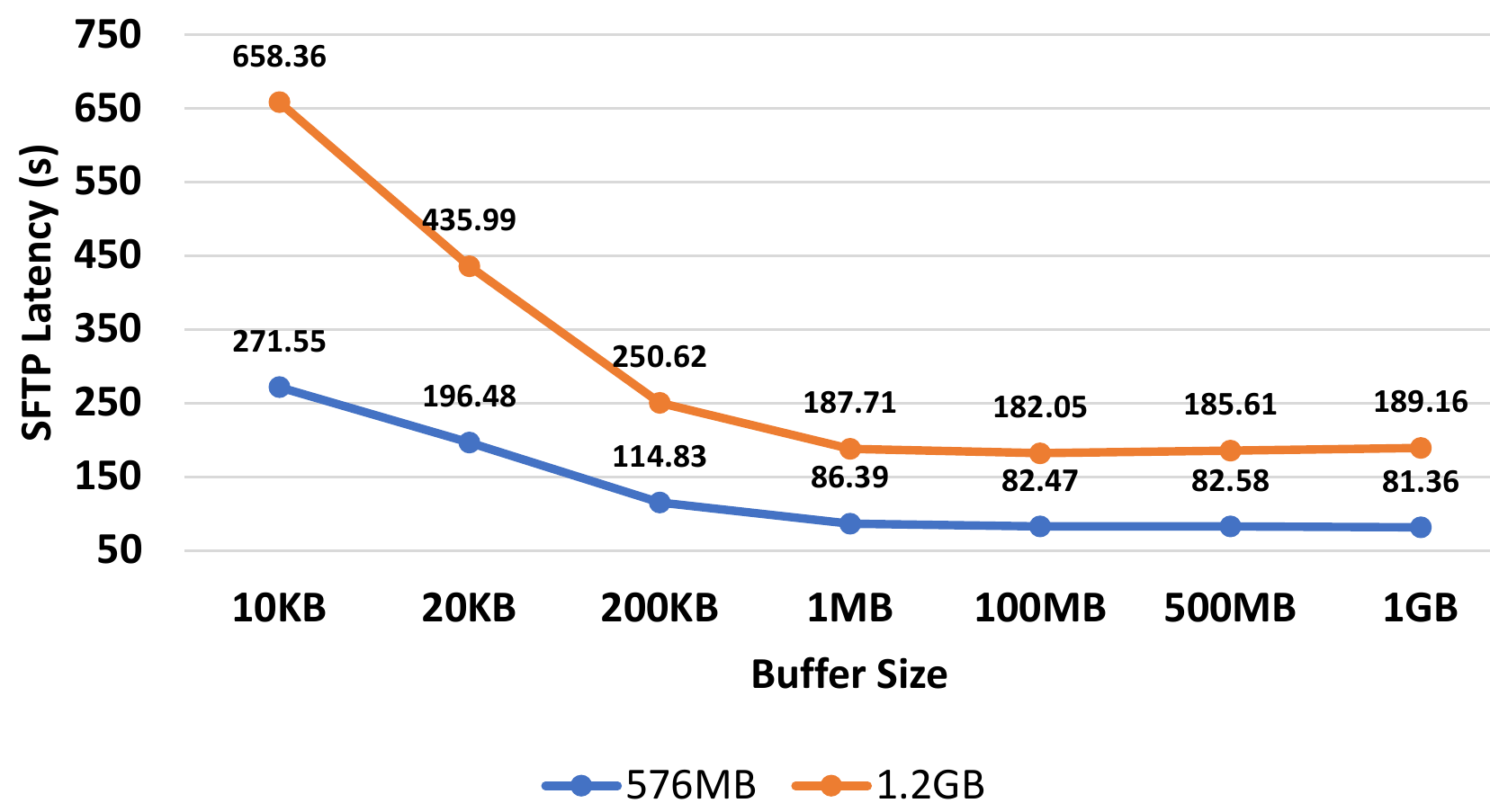}
\caption{\textit{SFTP} latency versus buffer size
}
\label{fig::SFTP_buffer} 
\end{figure}

We use \textit{Golang}'s \textit{SFTP} package to perform file transfer and find that the sender's buffer size parameter affects speed and latency of \textit{SFTP}. The latency refers to how long it takes to finish a particular operation. To find an optimal buffer size, we conduct \textit{SFTP} with different buffer sizes between two docker containers on physical Computer 1 and Computer 2 using a 576MB file and a 1.2GB file. Fig. \ref{fig::SFTP_buffer} shows how the buffer size affects the file transfer latency of \textit{SFTP}.
It can be observed that a buffer size of more than 1MB will not reduce the latency further. So we set the buffer size as 1MB in our system.

We use files of different types and various sizes to evaluate BOSS' feasibility and performance. Table \ref{table::file_list} lists those test files. Recall that a file sharing session involves four transactions as shown in Fig. \ref{fig::Module2}. We evaluate the latency of different transactions and the latency of the whole session. We evaluate how the file size and the parallel file transfers affect the latency. 

\begin{table}[h]
\centering
\caption{Test File List}
\begin{tabular}{lll}
\hline
Type & Size  & Describtion                       \\ \hline
.pdf & 67MB  & ``C++ Primer Plus" eBook          \\
.mp4 & 218MB & An over 5 hours song list video   \\
.tif & 576MB & The image of Moon from NASA       \\
.zip & 1.2GB & A collection of medical images    \\
.rar & 2.6GB & The compressed file of one movie  \\
.zip & 5.3GB & The compressed file of two movies \\ \hline
\end{tabular}
\label{table::file_list}
\end{table}

\begin{figure*}[!ht]
\begin{minipage}[c]{0.65\columnwidth}
\centering
\includegraphics[height=0.72\textwidth]{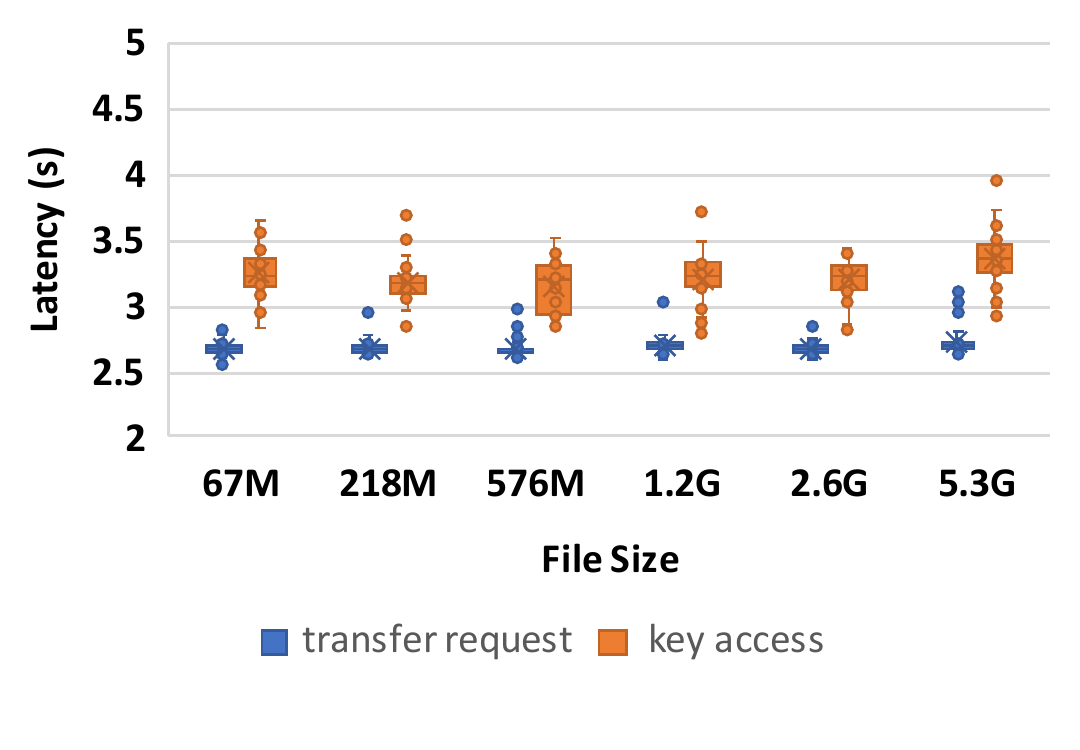}
\caption{Transaction latency of ``Tx: transfer request" and ``Tx: key access"}
\label{fig::1par_req} 
\end{minipage}
\hspace{0.3cm}
\begin{minipage}[c]{0.65\columnwidth}
\centering
\includegraphics[height=0.72\textwidth]{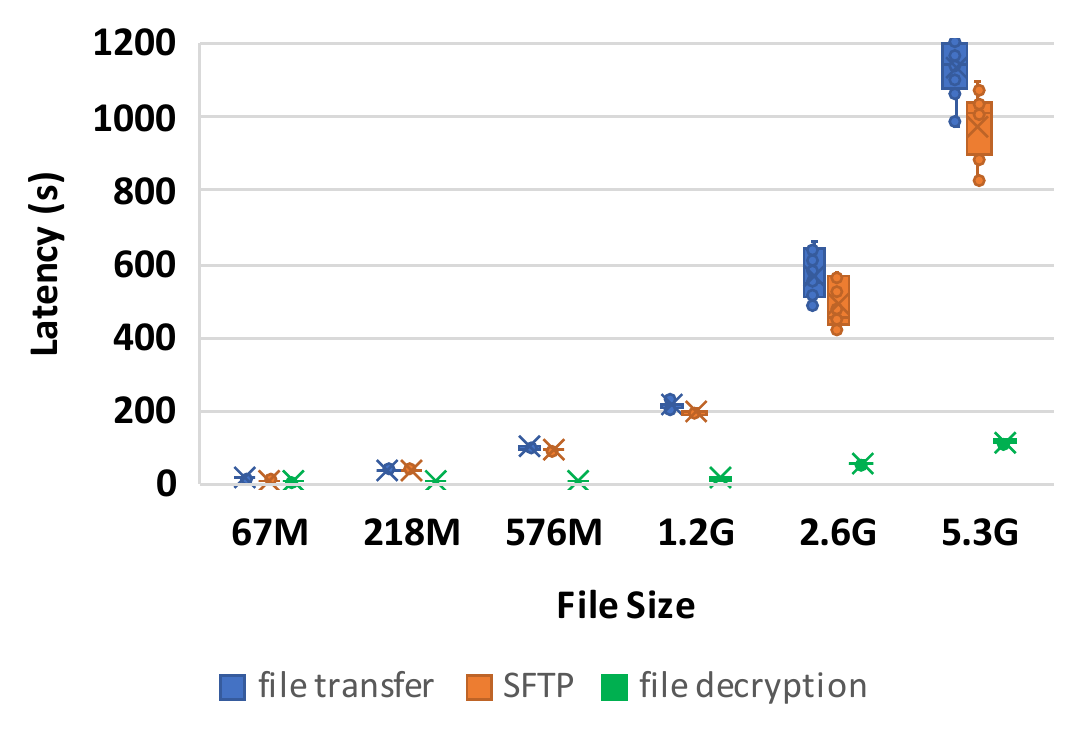}
\caption{SFTP latency and transaction Latency of ``Tx: file transfer" and ``Tx: file decryption"}
\label{fig::1par_tran_dec}
\end{minipage}
\hspace{0.3cm}
\begin{minipage}[c]{0.65\columnwidth}
\centering
\includegraphics[height=0.72\textwidth]{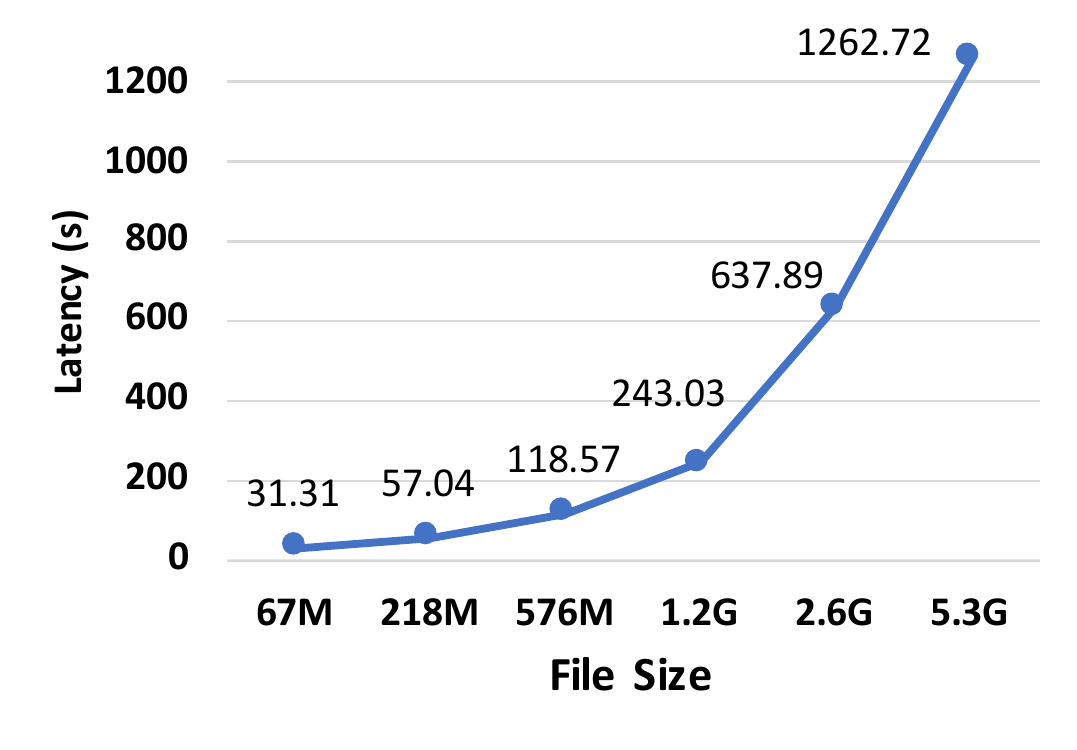}
\caption{File sharing session latency}
\label{fig::1par_whole}
\end{minipage}
\end{figure*}

\begin{figure*}[!ht]
\begin{minipage}[c]{0.65\columnwidth}
\centering
\includegraphics[height=0.64\textwidth]{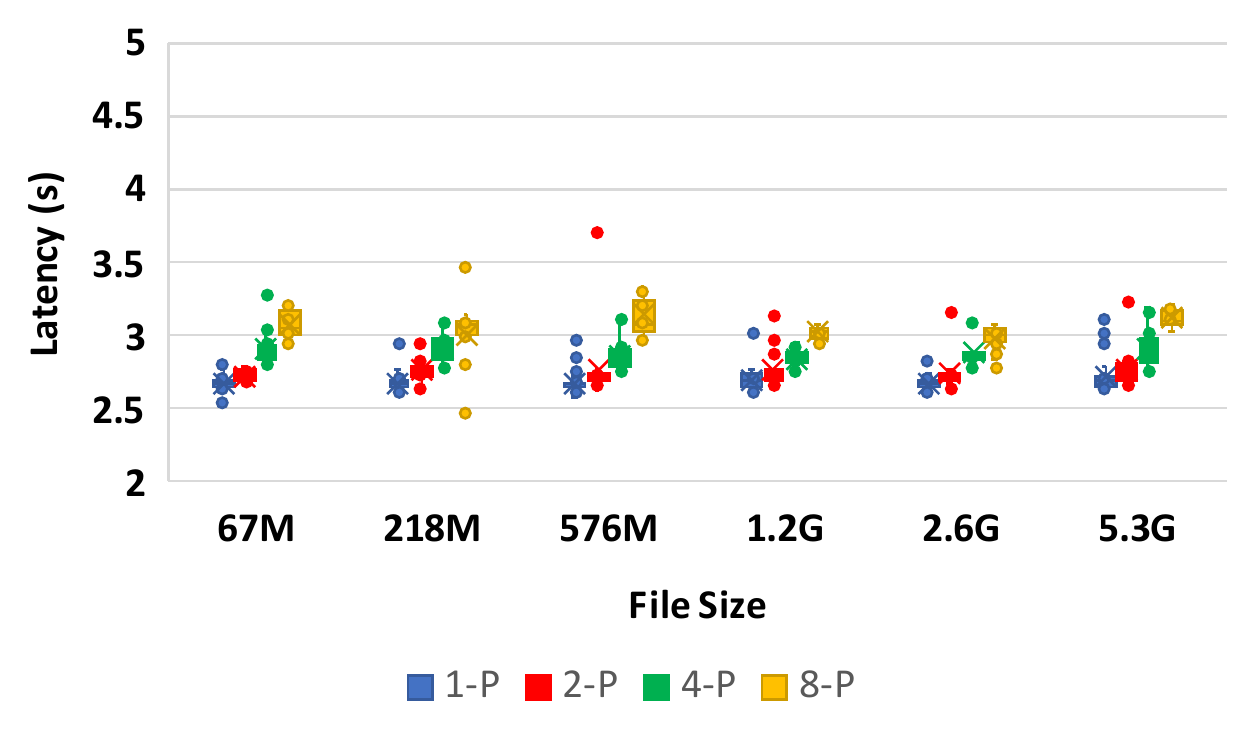}
\caption{Transaction latency of ``Tx: transfer request" in different parallel transfer cases}
\label{fig::mul_par_requestAgree_Tx} 
\end{minipage}
\hspace{0.3cm}
\begin{minipage}[c]{0.65\columnwidth}
\centering
\includegraphics[height=0.64\textwidth]{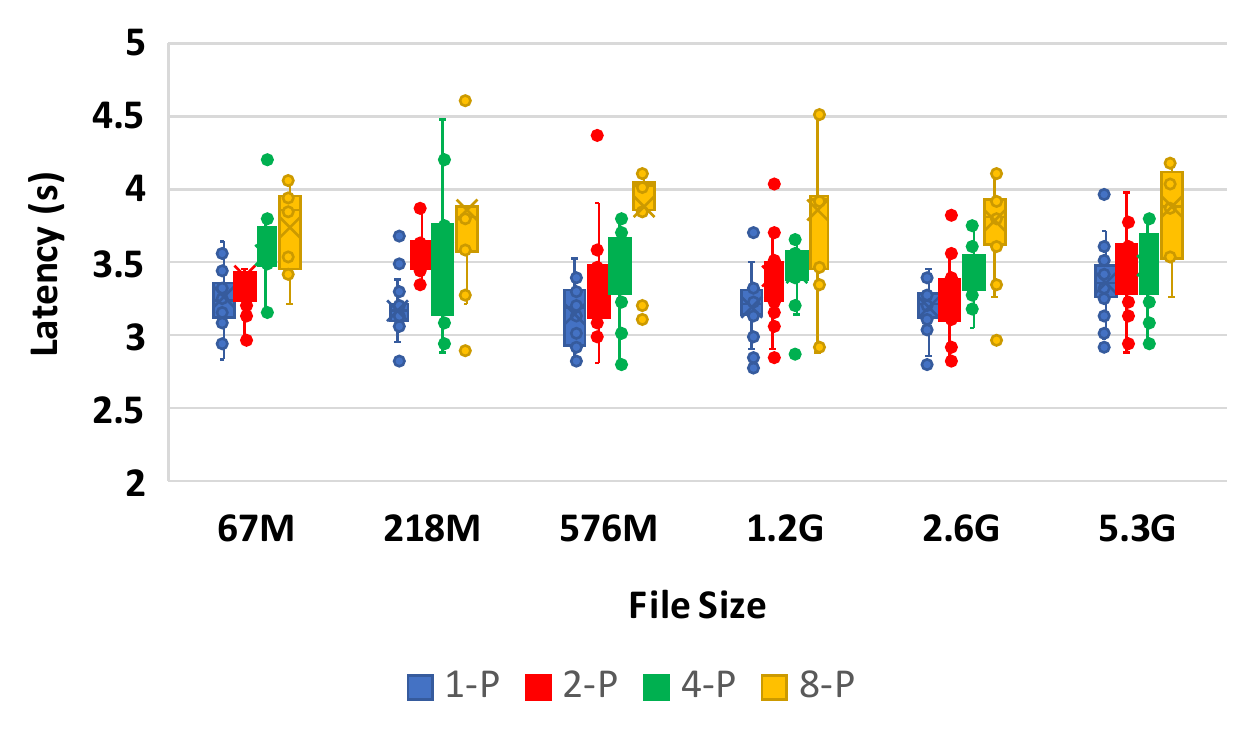}
\caption{Transaction latency of ``Tx: key access" in different parallel transfer cases}
\label{fig::mul_par_requestKey_Tx} 
\end{minipage}
\hspace{0.3cm}
\begin{minipage}[c]{0.65\columnwidth}
\centering
\includegraphics[height=0.64\textwidth]{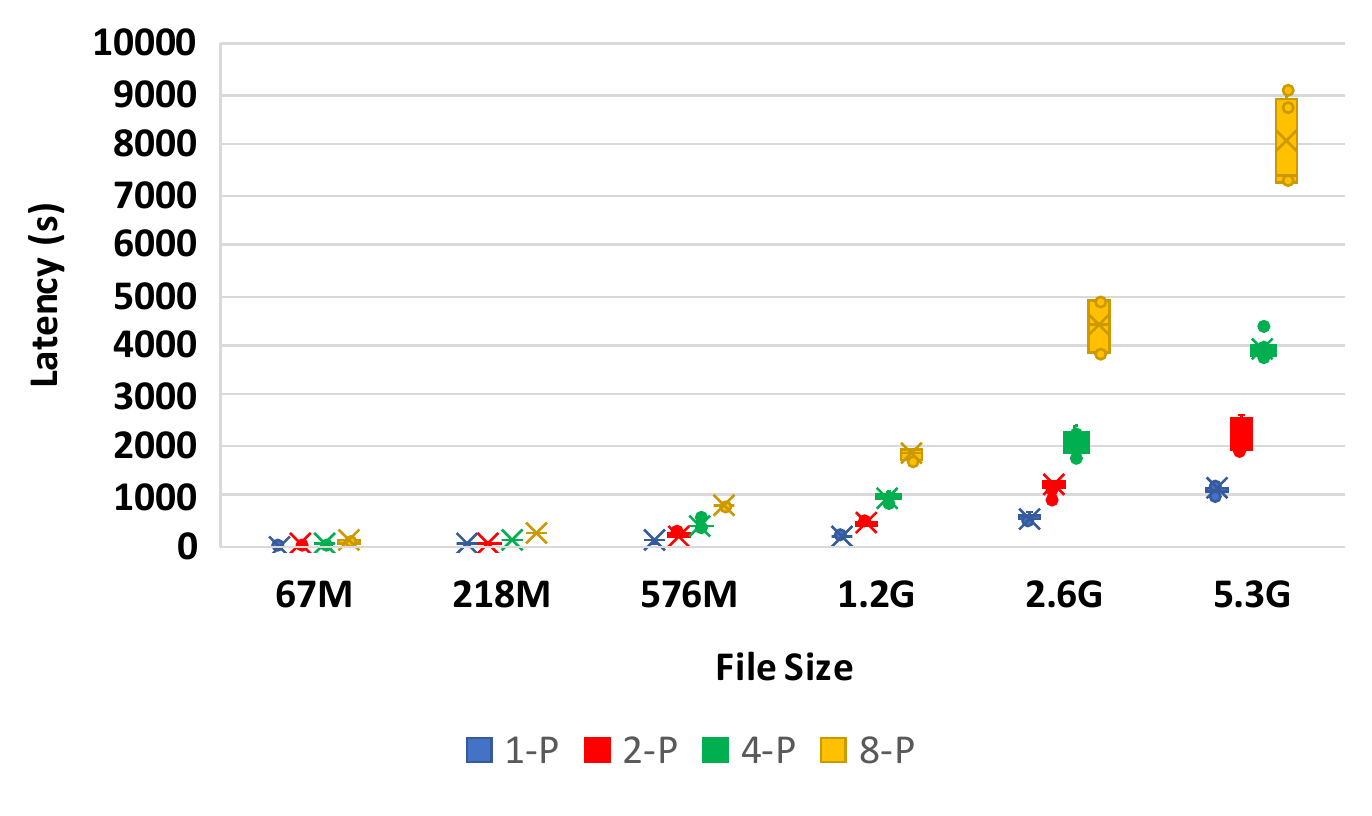}
\caption{Transaction latency of ``Tx: file transfer" in different parallel transfer cases}
\label{fig::mul_par_transfer_Tx}
\end{minipage}
\end{figure*}

\begin{figure*}[!ht]
\begin{minipage}[c]{0.65\columnwidth}
\centering
\includegraphics[height=0.64\textwidth]{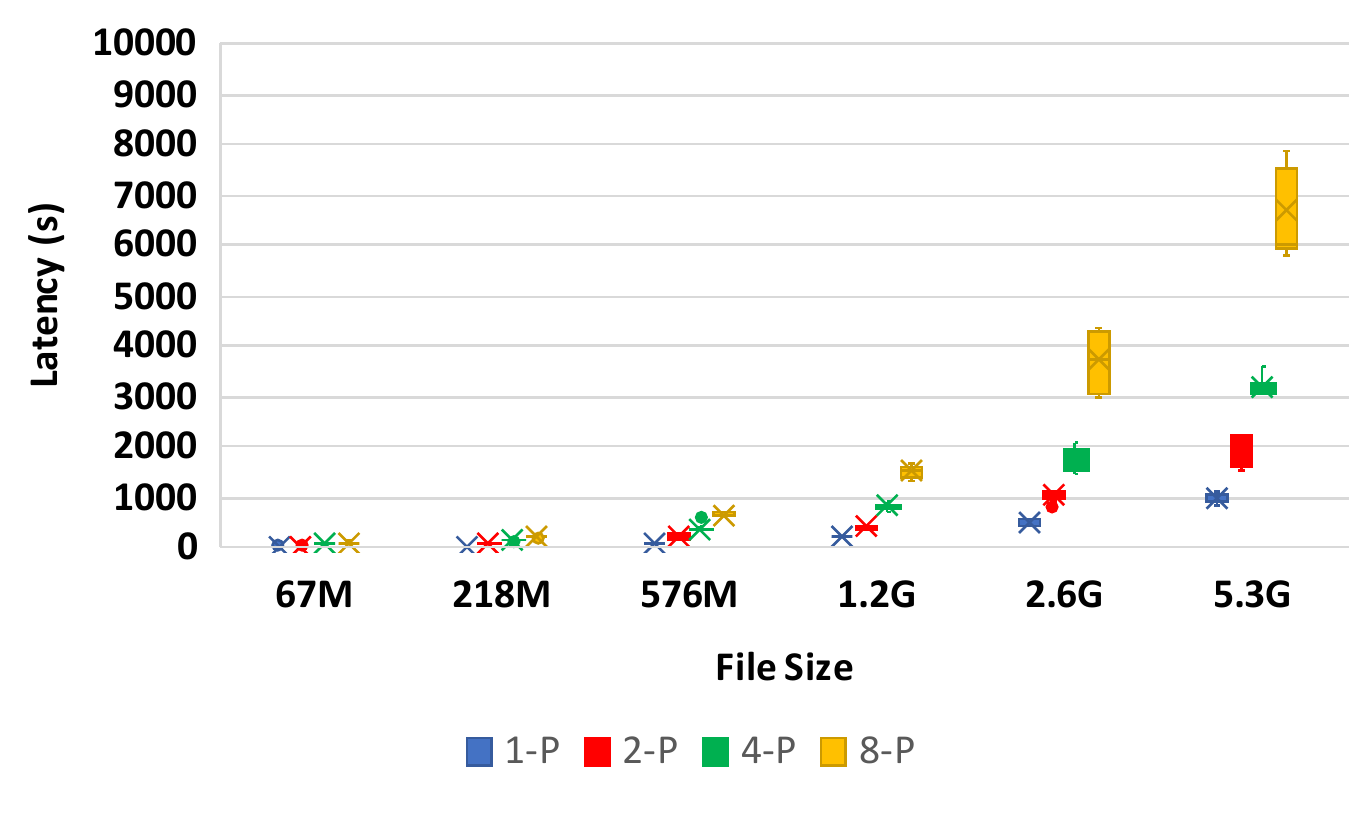}
\caption{SFTP latency in different parallel transfer cases}
\label{fig::mul_par_sftp} 
\end{minipage}
\hspace{0.3cm}
\begin{minipage}[c]{0.65\columnwidth}
\centering
\includegraphics[height=0.64\textwidth]{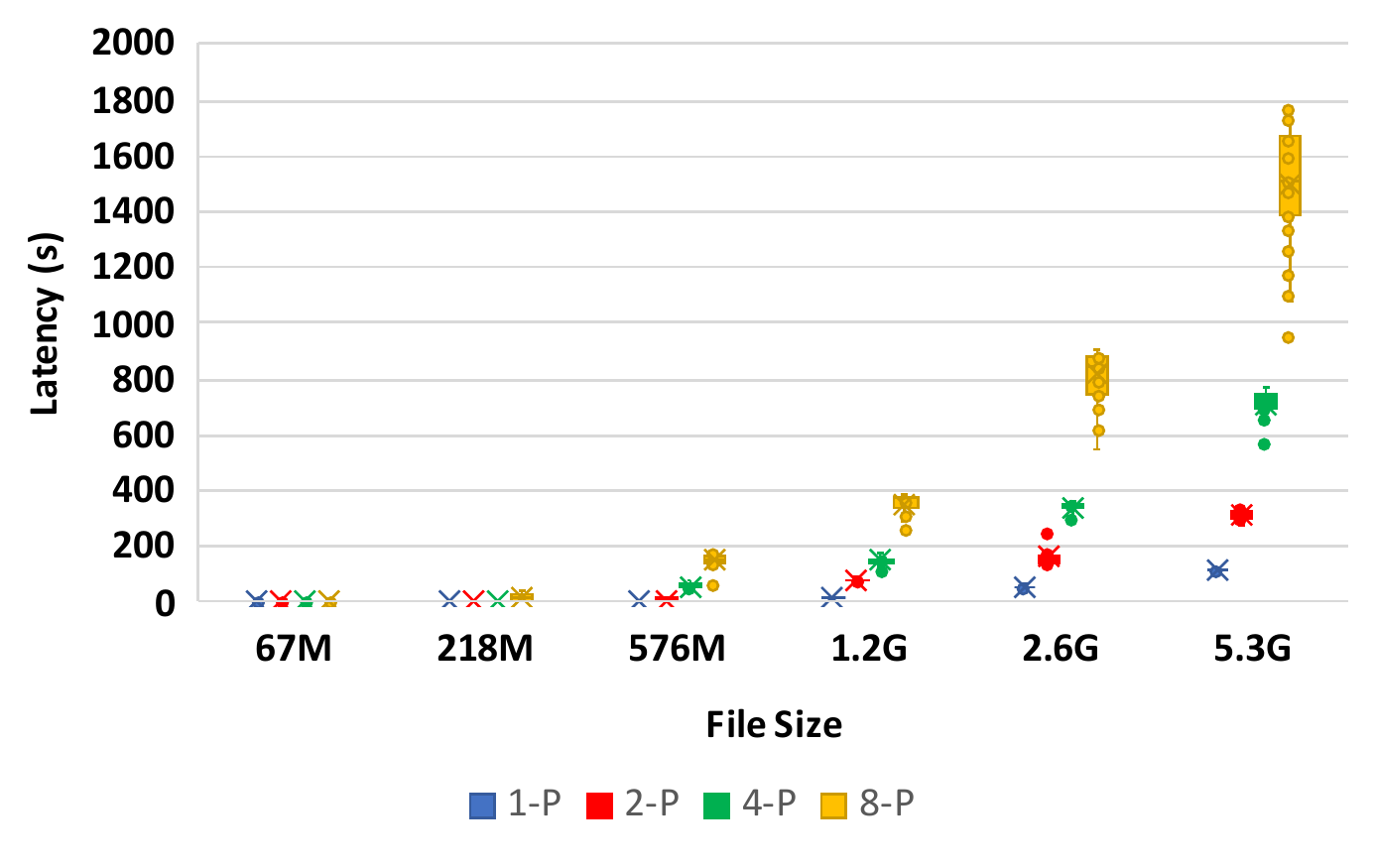}
\caption{Transaction latency of ``Tx: file decryption" in different parallel transfer cases}
\label{fig::mul_par_decrypt_Tx} 
\end{minipage}
\hspace{0.3cm}
\begin{minipage}[c]{0.65\columnwidth}
\centering
\includegraphics[height=0.68\textwidth]{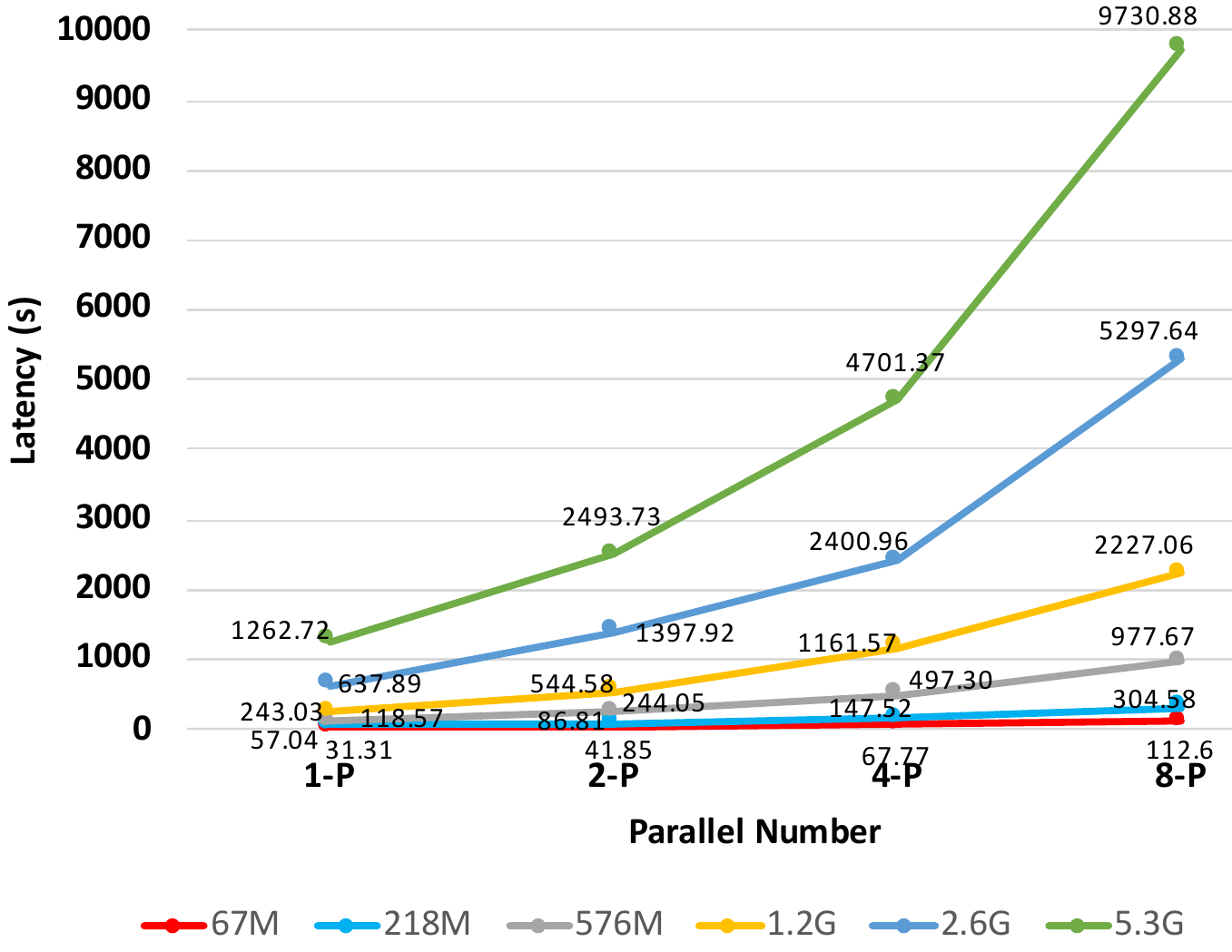}
\caption{Average file sharing session latency in different parallel transfer cases}
\label{fig::mul_par_ave_event} 
\end{minipage}
\hspace{0.3cm}
\end{figure*} 

\begin{table}
\centering
\caption{Percentage of SFTP Latency in ``Tx: file transfer" Latency}
\begin{tabular}{c|cccc}
\hline
Size                       & 1 Parallel & 2 Parallel & 4 Parallel & 8 Parallel \\ \hline
\multicolumn{1}{c|}{67MB}  & 75.30\%    & 85.33\%    & 93.38\%    & 92.01\%    \\
\multicolumn{1}{c|}{218MB} & 88.05\%    & 92.82\%    & 95.73\%    & 84.13\%    \\
\multicolumn{1}{c|}{576MB} & 92.43\%    & 94.86\%    & 87.50\%    & 82.33\%    \\
\multicolumn{1}{c|}{1.2GB} & 91.18\%    & 85.13\%    & 83.86\%    & 83.71\%    \\
\multicolumn{1}{c|}{2.6GB} & 86.24\%    & 85.71\%    & 84.36\%    & 84.08\%    \\
\multicolumn{1}{c|}{5.3GB} & 85.84\%    & 83.56\%    & 81.39\%    & 82.68\%    \\ \hline
\end{tabular}
\label{table::sftpPercentTran}
\end{table}

\subsubsection{\textbf{File Size}}
\label{sec::file_size_eva}

For each file in Table \ref{table::file_list}, we conduct the file transfer 32 times. Each time only one file transfer is performed. The results are shown in Figs. \ref{fig::1par_req}--\ref{fig::1par_whole}.

Fig. \ref{fig::1par_req} shows the latency of ``Tx: transfer request" and ``Tx: key access" across different file sizes. It can be observed that the latency of these two transactions are relatively stable and the file size has minor affect on the latency. This is reasonable since these two transactions only operate on the world state data.

Fig. \ref{fig::1par_tran_dec} shows the latency of ``Tx: file transfer" and ``Tx: file decryption". It can be observed that the latency of these two transactions rapidly increase as the file size increases.
The reason is ``Tx: file transfer" involves multiple cryptographic calculations to encrypt files and uses \textit{SFTP} to transfer the file. 
Fig. \ref{fig::1par_tran_dec} shows the latency of the \textit{SFTP} transmission. Table \ref{table::sftpPercentTran} shows the percentage of the \textit{SFTP} latency in the ``Tx: file transfer" latency. It can be observed that the \textit{SFTP} latency accounts for a large proportion of the whole ``Tx: file transfer" latency.  ``Tx: file decryption" involves multiple cryptographic calculation to decrypt the file. So it is reasonable that its latency is affected by the file size.

Fig. \ref{fig::1par_whole} shows the average latency of one file sharing session. It can be observed that the latency of one file sharing session rapidly increases as the file size increases. It is reasonable because normally more time is needed to transfer larger files.

\subsubsection{\textbf{Parallel Transfer}}
\label{sec::par_eva}

We also want to find out if we can transfer multiple files in parallel between two nodes and evaluate the performance.   
In our experiments, for each file, the number of simultaneous parallel file transfer sessions is 1, 2, 4 and 8, denoted as 1-P, 2-P, 4-P and 8-P in Figs. \ref{fig::mul_par_requestAgree_Tx}--\ref{fig::mul_par_ave_event}, which show the experiment results.
In each case, the file is transferred 32 times in total. For example, in the case of 4-P (4 parallel file transfer session), we perform the experiments 8 times and the file is transferred 32 times ($4\times8$).

We make the following observations. As the number of parallel sessions increases, the latency of ``Tx: transfer request" and ``Tx: key access" is relatively stable as shown in Fig. \ref{fig::mul_par_requestAgree_Tx} and Fig. \ref{fig::mul_par_requestKey_Tx}. However the latency of ``Tx: file transfer" and ``Tx: file decryption" increases as the number of parallel sessions increases. Fig. \ref{fig::mul_par_sftp} further shows the latency of \textit{SFTP} involved in ``Tx: file transfer" in different parallel transfer cases. Table \ref{table::sftpPercentTran} shows the percentage of \textit{SFTP} incurred latency in the ``Tx: file transfer" latency. It shows the same trend as ``Tx: file transfer" and accounts for the most of the transaction latency. \looseness=-1

Fig. \ref{fig::mul_par_ave_event} shows the average latency of one file sharing session in different parallel transfer cases. It can be observed that the average file transfer session latency increases as the number of parallel transfer sessions increases. This is reasonable because simultaneous parallel file transfer sessions share the network bandwidth. More sessions mean less network bandwidth for each file transfer session.
\section{Discussion}
\label{sec::discussion}

In this section, we discuss the impact of chaincode execution timeout on big file sharing, potential implementation of BOSS with other blockchain frameworks, difference between off-state and off-chain and out-of-band distribution of received files by users. \looseness=-1

\subsection{Chaincode Execution Timeout}


In Fabric, chaincode has a system wide execution timeout setting. The default timeout value is $30s$. In BOSS, the chaincode invokes SFTP for file transfer and waits for SFTP to finish. 
However, $30s$ is far from enough for transferring a big file and the chaincode will time out.
This timeout setting can be changed in the docker configuration file of Fabric, but cannot be changed at runtime based on the size of the file.
We find setting the timeout to infinity does not work in Fabric either and plan to revise Fabric as future work so that we can change the timeout flexibly.

\subsection{Implementation based on Other Blockchain Frameworks}

Bitcoin is not appropriate to implement BOSS. First, Bitcoin is a permissionless blockchain system. It is hard to implement access control. Second, Bitcoin essentially is a blockchain-based cryptocurrency application. There is no existing mechanism like smart contracts for programming. It will be a challenge and tedious to develop other applications over it.

Enterprise Ethereum \cite{enterpriseEthereum} can be used to implement a blockchain off-state sharing system.
Enterprise Ethereum \cite{Hyperledger_Besu}\cite{goquorum} is a permissioned blockchain framework for enterprise applications. It has many similar features to Fabric.
First, there is a private world state as well as a public world state for each enterprise Ethereum node. The private world state is only shared within a private group.
Enterprise Ethereum provides a private transaction manager, which can be used to keep transactions private among a private group. 
The private world state is similar to PDC in Fabric and can be used to manage the encryption key access required by BOSS.
Second, there is a private contract mechanism. A private contract is only available and executed on a subset of nodes. This is actually similar to customizable chaincode in Fabric and is flexible for the sender/receiver peers to implement their own business logic.
Third, the language of smart contracts is Solidity \cite{solidity_Wikipedia} in Ethereum. Solidity is a Turing-complete program language which can be used to implement a business logic. 

\subsection{Difference between Off-State and Off-Chain}

Off-state is different from off-chain \cite{poon2016bitcoin,coleman2018counterfactual,dziembowski2018general}, and existing off-chain protocols can not be utilized to ensure the safety of off-state data. 
The off-chain technique is designed to scale the blockchain systems and improve transaction throughput. 
It constructs transactions off-chains, and submits one transaction containing the final results to update the world state.
Essentially, the off-chain protocol is designed to guarantee the safety of the world state. 
The well-known off-chain technique is the payment channel in Bitcoin \cite{poon2016bitcoin}. A payment channel first locks some on-chain cryptocurrencies, i.e. the UTXOs state. Then it constructs transaction off-chains to conduct micropayments using locked cryptocurrencies, but does not commit these transactions involving micropayments to the blockchain. At last, it commits one transaction that represents final micropayment results to the main blockchain to update the UTXOs world state.
Off-chain can not solve the storage limitation issues or be used for big file sharing. 

\subsection{Out-of-band Data Distribution}

Once the file is transferred to a user through BOSS, the user may stealthily send the file to other parties through out-of-band commnunication but not through BOSS. It will be hard to trace those files. This is a generic problem in data sharing, not specific to our system. Potential solutions include digital watermark \cite{van1994digital} and traitor tracing \cite{boneh1999efficient}, which adds some additional information to the files. Another strategy will be to limit the user capabilities with the files to particular workstations and use physical security and particular configurations that allow file transfer only through BOSS.

\section{Related Work}
\label{sec::related_work}

Related work on blockchain-based data sharing systems does not focus on the big file sharing. It does not address the specific challenges in big file sharing or solve the problem.

\subsection{Data Sharing in Ledgers}
In most existing blockchain-based data sharing systems, shared data is stored in ledgers. Users set or get shared data through proposing transactions to the blockchain system. Transactions are the carriers of the shared data. 
Wang et al.\cite{wang2020blockchain} utilize the blockchain system as the back end to transmit and store data from the front-end wireless body area networks (WBAN). All blockchain nodes maintain the same data and all participants such as doctors and insurance providers share the same WBAN data.
Zhang et al. \cite{zhang2018towards} propose a blockchain-based secure and privacy-preserving personal health information (PHI) sharing scheme. The original PHI records are stored in the ledger of the private blockchain of each hospital, and the consortium blockchain keeps records of the indexes of the PHI. 
Jiang et al. \cite{jiang2018blochie} propose a blockchain-based platform for healthcare information exchange. They use two loosely coupled blockchains to handle different kinds of healthcare data, including highly sensitive electronic medical records (EMRs) and personal healthcare data (PHD). 
These existing schemes are not suitable for big file sharing applications because big files cannot be stored in blockchain ledgers due to the storage limitation as analyzed in Section \ref{sec::Storage_Limitations}.

\subsection{Data Sharing Outside Ledgers}
There are systems that store shared data outside the ledgers in blockchain systems. 
Xia et al. \cite{xia2017medshare} propose a blockchain-based solution to sharing medical data among cloud service providers. In their design, smart contracts retrieve requested data from existing database infrastructure, outside of a blockchain system. The retrieved results are recorded in the blockchain. It is impractical to record big files in a blockchain. 
Lu et al. \cite{lu2019blockchain} design a blockchain empowered privacy-preserved data sharing architecture for industrial IoT. They exploit the privacy-preserved federated learning algorithm to learn the data model and share the data model instead of revealing the actual data. They use the blockchain to retrieve and manage the accessibility of data.
Zhang et al. \cite{zhang2018fhirchain} choose to store and exchange only reference pointers of the sensitive data in the blockchain. A data requester should first get a token from the blockchain system and then use the token to access original data in the database outside the blockchain. 

These existing schemes do not focus on big file sharing. They also fail to consider security requirements such as dealing with dishonest users, and fail to design a protocol for secure off-state sharing.
Wang et al. \cite{PDCFabric} find misuse of PDC in Fabric may endanger a blockchain system and BOSS avoids such issues in the implementation.


\section{Conclusion}
\label{sec::Conclusion}

In this paper, we solve a novel problem---how to securely share big data files within a blockchain system and establish the chain of custody of the shared data.
Such a data sharing application is critical for protecting intellectual property (IP) theft and fighting industrial espionage in fields including biomedical research.
We denote data such as big files stored at a blockchain node but outside of the ledger as off-state.
Three challenges including storage space limitation, privacy requirement and security requirement are discovered in implementing a blockchain off-state sharing system (BOSS).
We carefully present our off-state sharing protocol. The transactions generated by our protocol will serve as auditing evidence for the chain of custody.
We implement BOSS over Hyperledger Fabric and conduct extensive experiments to evaluate the feasibility and performance of BOSS.





\bibliographystyle{IEEEtran}
\bibliography{file_sharing.bib}

\end{document}